\documentclass[acmlarge]{acmart}

\usepackage{graphicx}
\usepackage{subcaption}
\usepackage{array}
\usepackage[normalem]{ulem}
\usepackage[export]{adjustbox}
\usepackage{booktabs}

\usepackage{gensymb}

\newcolumntype{L}{>{\centering\arraybackslash}m{3cm}}
\newcolumntype{P}{>{\centering\arraybackslash}m{0.7\linewidth}}
\newcolumntype{R}{>{\raggedleft\arraybackslash}X}

\AtBeginDocument{%
  \providecommand\BibTeX{{%
    \normalfont B\kern-0.5em{\scshape i\kern-0.25em b}\kern-0.8em\TeX}}}

\setcopyright{rightsretained}
\acmJournal{IMWUT}
\acmYear{2023}
\acmVolume{7}
\acmNumber{1}
\acmArticle{14}
\acmMonth{3}
\acmPrice{}
\acmDOI{10.1145/3580820}

\begin{document}


\title[Feeling the Temperature of the Room: Unobtrusive Thermal Display of Engagement during Group Communication]{Feeling the Temperature of the Room: Unobtrusive Thermal Display of Engagement during Group Communication}

\begin{teaserfigure}
\centering
    \includegraphics[width=\linewidth]{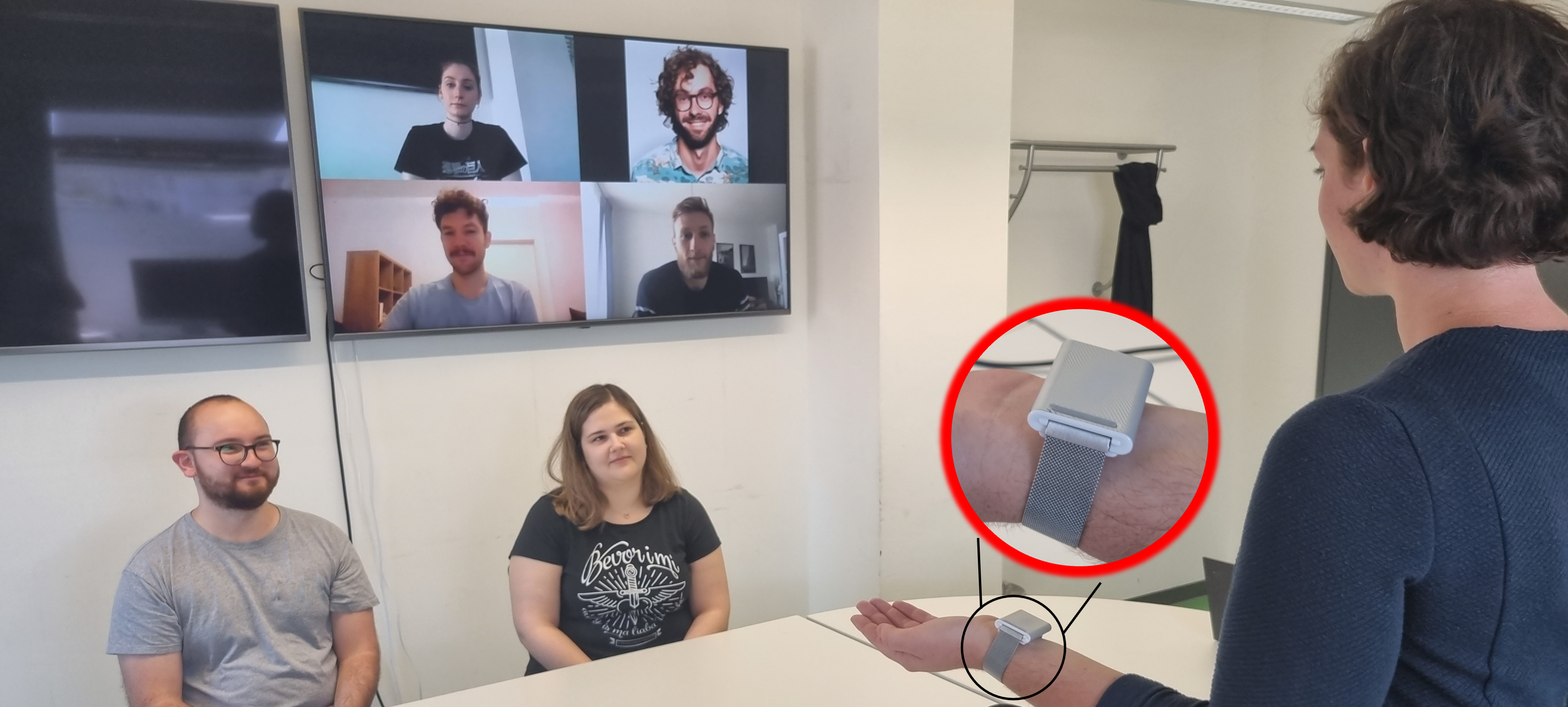}
\Description{An image showing a participant wearing the thermal bracelet, presenting to a hybrid audience with two in-person audience members and four virtual.}
\caption{We investigated thermal feedback to communicate audience engagement in a hybrid meeting scenario and found that thermal feedback can help a presenter to be more in tune with their audience without increasing workload. Faces are blurred for anonymization purposes.}
\label{fig:teaser}
\end{teaserfigure}

\author{Luke Haliburton}
\email{luke.haliburton@ifi.lmu.de}
\orcid{0000-0002-5654-2453}
\affiliation{%
  \institution{LMU Munich}
  \streetaddress{Frauenlobstr. 7a}
  \city{Munich}
  \country{Germany}
  \postcode{80337}
}

\author{Svenja Yvonne Sch\"{o}tt}
\orcid{0000-0003-1281-0230}
\affiliation{%
  \institution{LMU Munich}
  \city{Munich}
  \country{Germany}
}

\author{Linda Hirsch}
\orcid{0000-0001-7239-7084}
\affiliation{%
\institution{LMU Munich}
\city{Munich}
\country{Germany}
}

\author{Robin Welsch}
\orcid{0000-0002-7255-7890}
\affiliation{%
  \institution{Aalto University}
  \city{Espoo}
  \country{Finland}
}

\author{Albrecht Schmidt}
\orcid{0000-0003-3890-1990}
\affiliation{%
\institution{LMU Munich}
\city{Munich}
\country{Germany}
}

\renewcommand{\shortauthors}{Haliburton et al.}
\begin{abstract}
   Thermal signals have been explored in HCI for emotion-elicitation and enhancing two-person communication, showing that temperature invokes social and emotional signals in individuals. Yet, extending these findings to group communication is missing. We investigated how thermal signals can be used to communicate group affective states in a hybrid meeting scenario to help people feel connected over a distance. We conducted a lab study ($N$=20 participants) and explored wrist-worn thermal feedback to communicate audience emotions. Our results show that thermal feedback is an effective method of conveying audience engagement without increasing workload and can help a presenter feel more in tune with the audience. We outline design implications for real-world wearable social thermal feedback systems for both virtual and in-person communication that support group affect communication and social connectedness. Thermal feedback has the potential to connect people across distances and facilitate more effective and dynamic communication in multiple contexts.
\end{abstract}



\maketitle

\section{Introduction}
``Feeling the temperature of the room'' is a common metaphor in western languages that relates a group of people's mood to a temperature scale, because temperature is linked to socio-emotional experiences. In line with this, psychological research has shown that physical warmth promotes feelings of closeness to others~\cite{ijzerman_perceptual_2014} and communicates presence~\cite{wilson_hot_2016}, while cold induces feelings of separation~\cite{ijzerman_thermometer_2009} and absence~\cite{wilson_hot_2016}. Thus, temperature can be mapped to social emotions~\cite{ijzerman_cold-blooded_2012} and judgments of others~\cite{asch_forming_1961, ijzerman_temperature_2010}. 

Online communication is becoming increasingly ubiquitous, yet it omits many social and non-verbal aspects of communication~\cite{saatci_hybrid_2019}. As a consequence, recent work in HCI has explored embodied and tangible designs to connect remote communication partners~\cite{van_dijk_zooming_2022}. Similarly, recent works in Empathic Computing have attempted to share emotions to foster social connections between people~\cite{kim_sharing_2019, piumsomboon_empathic_2017, reichherzer_bringing_2021}. In this context, socio-thermal mappings have been applied to enhance the communication of emotions. Prior work used temperature, for example, to send asynchronous emotional signals~\cite{fujita_lovelet_2004,lee_thermo-message_2010} and to augment text messages~\cite{tewell_augmenting_2018}. To date, research has been limited to the case of sharing emotions between two users. We explore how thermal signals can be used to enhance real-time, spatially distributed communication with a group, extending what was once a one-to-one communication method to an n-to-one system.

In this paper, we explore thermal feedback for connecting groups across a distance by communicating group affect within the context of unobtrusive audience-speaker communication. We conducted a lab study ($N$=20) in which participants were given a series of personalized warming and cooling thermal stimuli representing audience engagement to the wrist while giving a presentation in front of a simulated hybrid audience. We collected data using questionnaires, an eye tracker, and semi-structured interviews. With this study, we aim to address the research question: \textit{To what extent can thermal signals be used to communicate group affect and increase feelings of social connectedness over a distance?}

Our results show that the socio-thermal signals are easy to understand and do not increase workload during presentations. We also found that thermal signals can promote audience awareness in the presenter and creates a correlation between social connectedness and presentation quality. However, the suitability of thermal signals varied for (dis)-engagement levels of the audience. Participants found thermal signals that show engagement encouraging and useful, while negative feedback (e.g., cold) was deemed distracting.

Overall, this paper contributes an exploration of thermal signals to communicate group affect in a hybrid presentation scenario and design recommendations for future design and research projects that use thermal signals for group communication. In particular, we derive seven design recommendations in total regarding group affect communication, social connectedness, and the implementation of a real-world system. This work extends the design space of thermal feedback to group communication and also adds a new modality to the space of audience-speaker communication. Our findings indicate that thermal feedback has the potential to facilitate more effective and dynamic communication and enhance social connectedness in both in-person and virtual contexts.

\section{Related Work}
In this section, we first present prior work on socio-thermal mapping and look at how thermal feedback has been explored in Human Computer Interaction (HCI). Next, we introduce relevant work on presentations and the communication of audience engagement. Finally, we provide a brief overview of unobtrusive interfaces and implicit interaction. 

\subsection{Socio-Thermal Mapping}
Our social relations, i.e., interpersonal feelings and associations between two or more people, can be expressed on a temperature scale. Warmth indicates a reduction of psychological distance (e.g., ``warming up to someone'') and coldness to an increase (e.g., ``being cold towards someone'')~\cite{ijzerman_perceptual_2014}. Indeed, the concept of warmth, along with competency, has been used effectively in social psychology to describe impressions of others~\cite{asbrock_stereotypes_2010,brambilla_effects_2010}. \citet{asch_forming_1961} showed in that a warm person is construed to be ``sociable'', ``popular'', and ``happy'', while opposing qualities, such as ``ruthlessness'' or ``unpopular'' refer to a cold person. This can be understood in terms of embodied realism, where direct perceptual experiences, temperature perception in this case, are tied to more abstract concepts, such as impression formation~\cite{lakoff_philosophy_1999}. Therefore, temperature can be seen as a readily accessible metaphor for the way we perceive and judge the affective state of others.

Research on embodied perception~\cite{niedenthal_embodiment_2005} indicates that the expression of a socio-thermal relation can be taken literally~\cite{ijzerman_thermometer_2009}. On the one hand, social exclusion is associated with a reduction of skin temperature~\cite{ijzerman_cold-blooded_2012} and estimates of room temperature~\cite{zhong_cold_2008}. On the other hand, physical warmth prompts favorable evaluations of others~\cite{williams_experiencing_2008} and closeness~\cite{ijzerman_temperature_2010}. The socio-thermal mapping has also been found to be contagious; participants who watch others experience cold or warm temperatures experience the sensation themselves~\cite{cooper_you_2014}. Thus, the socio-thermal relation can be considered a tightly coupled and bi-directional link that affects perception, motivation, physiology, and behavior.

In our work, we directly apply the socio-thermal mapping by having participants map an engaged or disengaged audience to a hot-cold scale. Prior research suggests that warmth equates to physical closeness, so we explore using thermal sensations to provide users with an indication of the attentiveness of an audience.

\subsection{Thermal Feedback in HCI}
HCI researchers (e.g.,~\cite{wilson_heat_2015, wilson_hot_2016, umair_exploring_2021}) have used temperature as a signal to understand elicited emotions and communicate information. There are several studies by~\citet{wilson_heat_2015, wilson_hot_2016} exploring the relationship between temperature and elicited emotions. They found a distribution of arousal and valence values assigned to various temperature inputs~\cite{wilson_hot_2016} and reported that most participants related warmth to presence and high quality of content while cold meant absence and poor quality of content~\cite{wilson_heat_2015}. A recent study by Umair and colleagues~\cite{umair_exploring_2021} asked participants to choose a thermal signal to receive when they were stressed, and half the participants chose cold while the other half chose warm. Umair recommends that individual personalization is crucial to making temperature signals understandable. In our work, we draw on the thermal associations of Wilson~\cite{wilson_heat_2015, wilson_hot_2016}, e.g., warmth is happy, satisfied, and excited while cold is tired, bored, and annoyed, but also allow for personalization of the direction mapping as recommended by Umair~\cite{umair_exploring_2021} to ensure that the temperature signaling of affective states are understandable for all participants.

Previous work has identified locations on the body that are best suited to thermal feedback. Several groups have compared temperature sensitivity and device usability on multiple body parts~\cite{maeda_thermodule:_2019,kappers_thermal_2019,wilson_like_2011}. In general, hand~\cite{zhu_sense_2019,wilson_like_2011} and wrist-worn~\cite{maeda_thermodule:_2019,kappers_thermal_2019,wilson_like_2011, song_hot_2015} devices were found to be effective and usable and have therefore been used in a range of application scenarios. Researchers have used thermal signals to communicate non-verbal messages to a partner~\cite{fujita_lovelet_2004,lee_thermo-message_2010}, to communicate color to visually impaired people~\cite{bartolome_exploring_2020}, as mobile icons~\cite{wilson_thermal_2012}, and as a mobile display~\cite{wettach_thermal_2007}. In a more abstract approach, \citet{poguntke_rainsense_2018} used thermal feedback to communicate weather information and stress~\cite{poguntke_designing_2019}. In general, researchers have found that temperature can be a useful modality but it is not suitable for communicating complex (i.e. high-bandwidth) messages~\cite{kappers_thermal_2019}. Rather than using temperature on its own, several groups have also used it to augment other forms of communication. \citet{tewell_heat_2017} used temperature to add arousal information to text messages, while \citet{akazue_using_2017} augmented images. \citet{el_ali_thermalwear_2020} used temperature to add arousal to voice messages for individuals with emotional prosody impairments. In all cases, temperature was an effective method of adding emotional signal to standard forms of communication.

\subsection{Augmenting Presentations and Communicating Engagement}
Humans normally gauge how interested an audience is through a combination of explicit (e.g., applause) and implicit non-verbal~\cite{knapp_nonverbal_2013} (e.g., eye contact and posture) signals. However, with large or virtual audiences, non-verbal cues are often lacking~\cite{saatci_hybrid_2019}, so researchers have explored automatic systems for communicating audience feedback. Most of the prior work in the field has focused on measuring the affective state of the audience. Researchers have used explicit feedback methods such as the \textit{Live Interest Meter} by Rivera-Pelayo~\cite{rivera-pelayo_live_2013}, where the presenter periodically polls the audience for feedback. HCI researchers have also explored the extent to which technology can be used to implicitly measure engagement~\cite{stevens_methods_2007}. Numerous physiological measurements have been used to quantify engagement, such as electroencephalography (EEG)~\cite{zhang_correlating_2014}, accelerometers~\cite{martella_how_2015,englebienne_mining_2012}, cameras~\cite{fujii_sync_2018,hernandez_measuring_2013}, electrodermal activity (EDA)~\cite{garbarino_empatica_2014}, galvanic skin response (GSR)~\cite{latulipe_love_2011}, and heart rate~\cite{atroszko_cardiovascular_2014}. Heart rate~\cite{konvalinka_synchronized_2011}, EDA~\cite{jaimovich_contagion_2010, gashi_using_2019} and breathing rate~\cite{bachrach_audience_2015} have been used to measure synchrony (i.e., the development of interdependent physiological states) between presenter and audience.

Despite the plethora of work that focused on measuring engagement and synchrony, very few researchers have addressed the issue of communicating this information to the presenter. Hassib and colleagues~\cite{hassib_design_2018} outlined a design space for audience sensing and feedback with dimensions: sender and receiver cardinality (i.e. the number of people being sensed or receiving feedback), location of audience (collocated or distributed), feedback synchronicity (synchronous or asynchronous), sensor location (on-body or environment), and feedback sensing style (implicit or implicit). As an extension of this work, they developed the \textit{EngageMeter}~\cite{hassib_engagemeter:_2017}, which uses EEG to measure audience engagement implicitly and provide both live and post-hoc visual feedback. Similarly, Murali et al.~\cite{murali_affectivespotlight_2021} implemented a visual spotlight system for virtual meetings that highlights audience affective states. In both cases, the assistive technology was perceived as useful and helped the presenter understand their audience~\cite{hassib_engagemeter:_2017, murali_affectivespotlight_2021}, but both use visual feedback, which has been shown to demand attention and be distracting~\cite{lavie_distracted_2005}.

To our knowledge, the usage of real-time non-visual feedback as a modality to communicate audience affective states to a presenter is yet to be explored. Given the high attentive demand of visual stimuli~\cite{lavie_distracted_2005} and the emotional metaphors associated with temperature~\cite{wilson_hot_2016}, thermal feedback is a promising modality for this use case.

\subsection{Unobtrusive Design}
Unobtrusive design enables the user to easily ignore an interface and to keep it in the periphery of their attention~\cite{bakker_design_2013, weiser_coming_1997}. Research about peripheral and implicit interactions aims to reduce stress~\cite{ledzinska_metaphorical_2017, kiss_stressed_2019} and avoid information overflow and interruptions~\cite{weiser_coming_1997, raudanjoki_perceptions_2020, kim_designing_2009, bakker_introduction_2016}. Thus, users can concentrate on their main task at hand~\cite{hausen_peripheral_2012, hausen_everyday_2014}.

Designing for peripheral interaction means designing for attention-inviting and non-demanding interfaces~\cite{bakker_introduction_2016}. A shift in attention can be triggered by, among other things, evoking positive or negative emotions, but these also require personalization~\cite{bakker_design_2013}. Thus, designing for implicit interaction means enabling secondary activities to run in the background of attention, often in parallel to explicit tasks~\cite{serim_explicating_2019,ju_design_2008}. Interactive wearables are also often considered unobtrusive~\cite{rekimoto_gesturewrist_2001, zheng_unobtrusive_2014, gashi_using_2019, lietz_wearable_2019, di_lascio_unobtrusive_2018} as long as they are adapted to the user's skin and temperature sensitivity and stay non-disruptive to movement~\cite{genaro_motti_overview_2015}, and prior work has explored novel affective wearable displays, such as chronometry~\cite{umair_towards_2019}. Costa et al.~\cite{costa_emotioncheck_2016} have shown that wearable affective feedback systems can be effectively implemented with minimal impact on attention. Researchers are increasingly studying emotions elicited by unobtrusive wearables using thermal feedback (c.f.,~\cite{gradl_overview_2019, nasser_thermo-haptic_2019, suen_development_2021}). Such interfaces, including the one used in our study, aim to not pose additional cognitive load by their unobtrusive design and provide a peripheral interaction that is easy to ignore if desired.

\section{Method} 
We conducted a within-subjects lab study to explore thermal feedback for communicating group affective information in the context of presentations. Each participant delivered two presentations to a hybrid audience based on one-page prompts. One presentation was delivered with thermal feedback and one without. The order was counter-balanced using a Latin Square~\cite{bradley_complete_1958}. This study design was chosen so that each participant would experience thermal feedback as an affect-communication modality in a controlled scenario, enabling us to investigate the technology in an exploratory manner.

\subsection{Apparatus}
All experiments took place in a conference room equipped with large wall-mounted screens and a meeting table. Each participant sat on one side of the table while the in-person audience sat on the opposite side, and the virtual audience was displayed on the screen on the wall opposite the presenter (see \autoref{fig:teaser}).

We used an Embr Wave\footnote{\url{https://embrlabs.com/}} thermal wristband to provide thermal feedback to the participants. The device has a resolution of approximately 0.1\degree C and can be adjusted with a rate of change of 0.1--1 \degree C/s~\cite{smith_augmenting_2017}. The Embr Wave device operates by selecting a temperature offset, rather than an objective temperature set point, and the device pulses to this temperature offset. This model has previously been used in HCI studies to provide a thermal signal~\cite{umair_exploring_2021, smith_augmenting_2017}.

We also equipped participants with a wearable Pupil Labs\footnote{\url{https://pupil-labs.com/}} Pupil Core eye tracker, which has commonly been used to capture unconstrained eye motion~\cite{kassner_pupil_2014, jungwirth_eyecontrol_2018, santini_art_2018}.

\subsection{Task}
The primary experimental task was for each participant to give two 3-minute presentations. Although 3 minutes is a relatively short presentation window in which to engage an audience, it is representative of multiple real life scenarios such as startup pitches in entrepreneurship (typically 1-5 minutes) or academic 3-minute-thesis formats. Since the aim of the task was for participants to give a presentation with very little information, we opted for this short but applicable timescale. The participants were given a one-page prompt for each presentation with information about a fundraiser\footnote{The prompts are included in the supplementary material.}. Additionally, we provided them with materials to take notes and as much time as they wished to prepare for the presentations. We asked them to imagine that they were raising money for each of the fundraisers and gave all participants the same two topics --- a new magazine to empower young girls and a newly invented automated toothbrush. The two topics were based on real, successful Kickstarter\footnote{\url{https://www.kickstarter.com/projects/amabrush/amabrush-worlds-first-automatic-toothbrush} and {https://www.kickstarter.com/projects/2110119100/kazoo-magazine}, last accessed \today} campaigns.

All participants presented to a hybrid audience consisting of four virtual and two in-person audience members (see \autoref{fig:teaser}). This setup was chosen because hybrid events are becoming increasingly ubiquitous and often lack non-verbal cues~\cite{saatci_hybrid_2019}. We recorded two reaction videos of the virtual audience before the study, which consisted of three real people and one ``camera off'' person. The virtual audience members were displayed as individual panels in ``gallery view'', as is common in many commercial video conferencing systems. While recording the videos, the virtual audience was instructed to act engaged or disengaged according to predefined timings. One author was present in the call as the ``camera off'' person and verbally communicated to the virtual audience members when it was time to act engaged or disengaged throughout the recording process. During the presentations, the reaction video was played without sound. The full timeline of both temperature and facial expressions is shown in \autoref{fig:timeline}. During the actual task, each participant was presented with the pre-recorded virtual audience videos while they delivered their presentations. Two authors were also present as in-person audience members and reacted according to the same script timing as the virtual audience. We chose to use audience facial reactions (without thermal feedback) as the control condition since this is the most common type of immediate feedback experienced during presentations.

The task, as designed, uses a Wizard of Oz approach since the feedback from the audience is simulated. This approach was chosen primarily to reduce the variability between the feedback each participant receives. Since this is the first investigation into thermal feedback as a mechanism to communicate group affect during presentations, it was most important to have each participant experience a range of audience feedback that stays consistent across participants. We aimed to investigate how users perceive both positive and negative feedback and how they could imagine integrating such feedback into their presentations.

\subsection{Experimental Procedure}
\begin{figure*}[ht] 
    \centering
    \includegraphics[width=\linewidth]{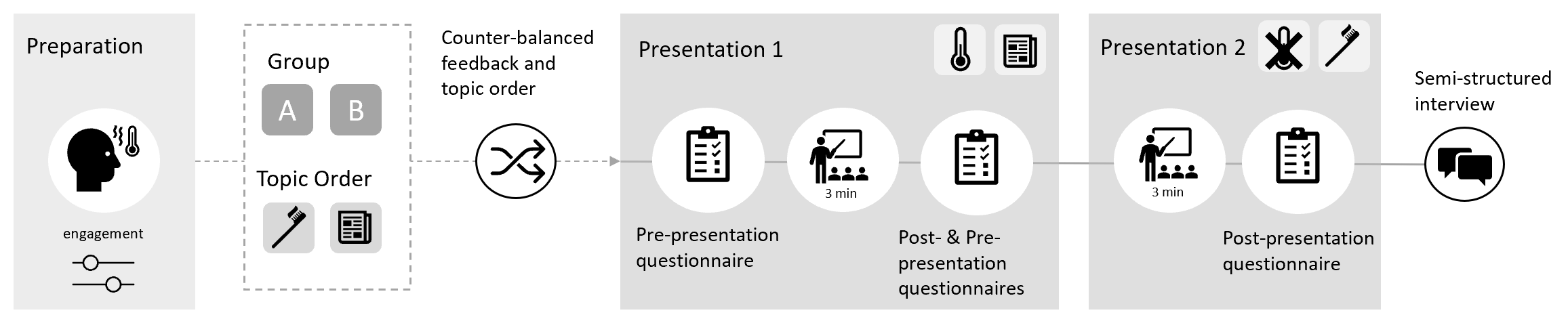}
    \caption{Overview of the experimental procedure. As a preparation, the thermal feedback is personalized to participants' sensation. The participants give one presentation with and one presentation without thermal feedback in a counter-balanced order.}
     \Description{Diagram showing the overview of the experimental protocol. As a preparation, the thermal feedback is personalized to participants' sensation. The participants give one presentation with and one presentation without thermal feedback in a counter-balanced order.}
     \label{fig:experimental_design}
\end{figure*}

The experiment consisted of two distinct phases. In the first phase, we personalized the socio-thermal mapping and calibrated the equipment for each participant. In the second phase, the participants held two presentations in front of a hybrid audience with the aim of raising funds based on real organizations --- a social magazine and a toothbrush company. The experimental protocol is shown in \autoref{fig:experimental_design}.

\subsubsection{Preparation Phase: Personalization and Calibration}
After introducing the study and obtaining informed consent, we equipped the participants with the Pupil Labs eye tracker. We first calibrated the eye tracker using the standard procedure outlined by the manufacturer and the associated \textit{Pupil Capture} software and let participants get accustomed to having the device present in their peripheral vision.

Following the eye tracker setup, we equipped the participants with an Embr Wave thermal wristband~\cite{umair_exploring_2021, smith_augmenting_2017}. The participants were instructed to place the wristband on whichever wrist they preferred, with the thermal module on the inner side of the wrist. We personalized the thermal feedback for each participant according to the recommendation by Umair~\cite{umair_exploring_2021}. The participants could chose a maximum threshold they were comfortable with, up to $\pm8\degree$C. The $\pm8\degree$C maximum range was based on common values used in prior work~\cite{wilson_heat_2015, umair_exploring_2021}. The participants were presented with the thermal signals in each direction in a counter-balanced order. The participants were then asked to choose one end of the scale (Completely Engaged to Completely Disengaged) to associate with each endpoint.

We constructed a linear 5-point mapping between temperature and engagement based on the end-points established in the personalization step. For example, if a participant was comfortable with the initial temperature range and associated +8$\degree$C with a Completely Engaged audience and -8$\degree$C with a Completely Disengaged audience, we would construct the mapping depicted in \autoref{tab:engagement_mapping}. To confirm the mapping, we presented participants with the thermal feedback again and asked them to recall the associated mapping. We gave the participants the opportunity to change their mapping if the signal did not comply with their expectations. The aim of this process was to ensure that participants knew the mapping between temperature and engagement level and understood the thermal signal before starting the main phase of the study. 

\begin{table}[ht]
    \caption{Example personalized mapping between temperatures and the engagement of an audience.}
    \label{tab:engagement_mapping}
    \begin{tabular}{cc}
        \toprule
        Engagement Level & Temperature\\
        \midrule
        Completely Disengaged & -8$\degree$C\\
        Disengaged & -4$\degree$C\\
        Neutral & 0$\degree$C\\
        Engaged & 4$\degree$C\\
        Completely Engaged & 8$\degree$C\\
    \bottomrule
\end{tabular}
\end{table}

Before moving on to the main experiment task, we measured the temperature of the room using a Bosch UniversalTemp infrared thermometer and asked the participants for their subjective perception of the room temperature. The average room temperature during experiments was 26.8$\degree$C (SD=1.1). None of the participants felt cold during the study. 35\% felt neutral, 50\% felt warm and 15\% felt hot.

\subsubsection{Main Phase: Presentations}
Following the personalization phase, every participant took part in both experimental conditions: (1) presenting with thermal feedback and (2) presenting without thermal feedback. Participants received facial reactions from the hybrid audience members in both conditions. The audience was introduced to the participants as a `simulated audience' to communicate that the reactions were pre-recorded and consequently they could not interact with the audience. This was communicated to prevent an interruption in flow for the scenario where participants might try to ask the audience to react in a specific way (e.g., raising their hands). The order of the conditions and the topics were counter-balanced using a Latin Square~\cite{bradley_complete_1958}, resulting in four different combinations of topics and feedback availability. Each participant delivered two total presentations, one with thermal feedback and one without.

\subsection{Measures}
We are investigating a novel feedback modality for presentations and therefore collected both quantitative and qualitative data from multiple sources to understand the user experience and the impact on the user. As such, we collected eye tracker data as a proxy for distraction, questionnaire responses for usability and social connection, and user interviews for deeper qualitative insights.

We used the \textit{Pupil Capture} software to collect data from the eye tracker. The eye tracker was used to detect glances at the thermal feedback device, which would indicate that the feedback drew the participant's attention. The thermal bracelet was controlled with a custom Python script, which also logged timestamps for the thermal signals. The thermal feedback followed the same script as the audience reactions. 

The participants completed questionnaires before and after each presentation to capture perceptions on performance, workload, usability, and connection. Before each presentation, the participants were asked about their expected performance: ``How well do you think you will do during your talk?''. Following each presentation, the participants responded to a questionnaire with standardized scales for workload (NASA-TLX~\cite{hart_development_1988,hart_nasa-task_2006}), usability (AttrakDiff~\cite{hassenzahl_attrakdiff_2003}), and presentation confidence (PRCS~\cite{hook_short-form_2008}), as well as subjective Social Connectedness and Perceived Presentation Quality questions from Murali's work on audience communication~\cite{murali_affectivespotlight_2021} and perceived stress~\cite{thomee_prevalence_2007}. We also collected basic demographics at the end of the final questionnaire.

As a final step, every participant completed a semi-structured interview to gather qualitative data. We had guided questions, such as asking participants about their feelings towards the thermal feedback, whether they found it distracting, how usable they found it, and how they would imagine using such a system in a real presentation.

\subsection{Analysis}
We collected data from questionnaires, interviews, and an eye tracker. Each of these data sources has a separate method of analysis.

We analyzed the standard scales (e.g. NASA-TLX~\cite{hart_development_1988,hart_nasa-task_2006}) according to their original documentation. We performed t-tests when normality was satisfied according to Shapiro-Wilk tests and performed paired Wilcoxon signed-rank tests with Bonferroni corrections on all other questionnaire responses. 

The audio from each semi-structured interview was recorded and transcribed verbatim. As a first step, the interview transcripts were coded using open coding by two authors. After a discussion, we identified themes in the responses and completed a second pass with a focus on specific questions from the interview protocol, such as whether the participants prefer the aggregated signal showing mean affect or some alternative. Three authors discussed and agreed on the final high-level themes.

For the eye tracker, we correlated fixation events with different thermal signal levels to analyze whether the bracelet was distracting. For our purpose, we define distraction as a shift in attention from the primary task, while disruption is defined as when the primary task is interrupted by a distraction~\cite{graydon_distraction_1989}. As such, if we record moments where the user's attention is drawn to the bracelet to the extent that they look at it, we would classify this as a disruption due to strong distraction. We recorded timestamps for the thermal signal provided by the Embr Wave bracelet and the facial feedback of the virtual audience. We checked these recordings to ensure that the signals matched our predefined script and used the timeline when analyzing the data from the eye tracker. The second author coded the eye tracker videos and labeled each time the participant glanced at the thermal bracelet. The initial analysis step was a manual control for device glances and fixations. One author looked at the recordings for all presentations with thermal feedback at two times speed using the Pupil Player v3.3.0 software. Within this software, eye gaze was visualized as a closed gaze circle of radius 20 and gaze movement was indicated using a gaze polyline of thickness 2. We used the fixation detector plugin, which calculates fixations based on the positional angle of the eyes, with default parameters for the dispersion threshold parameter, minimum duration, and maximum duration.

\subsection{Participants}
We recruited $N$=20 participants, aged 22-55, $M$=27, $SD$ = 6.97 (13 female, 7 male). Sixteen of the participants had a bachelors level of education or higher, one completed vocational training, and three have a high school diploma. Participants give an average of 1.5 presentations per week. Most of the participants had little to no experience with thermal feedback. The compensation was based on an hourly rate of 10 €/hour. The only prerequisite for participation was proficiency in English. The study was approved by the ethics committee within the University Faculty\footnote{Details removed for anonymization purposes}. Below, we will refer to participants with their study ID (e.g. P5).

\section{Results}
\begin{figure*}[ht] 
    \centering
    \includegraphics[width=0.75\linewidth]{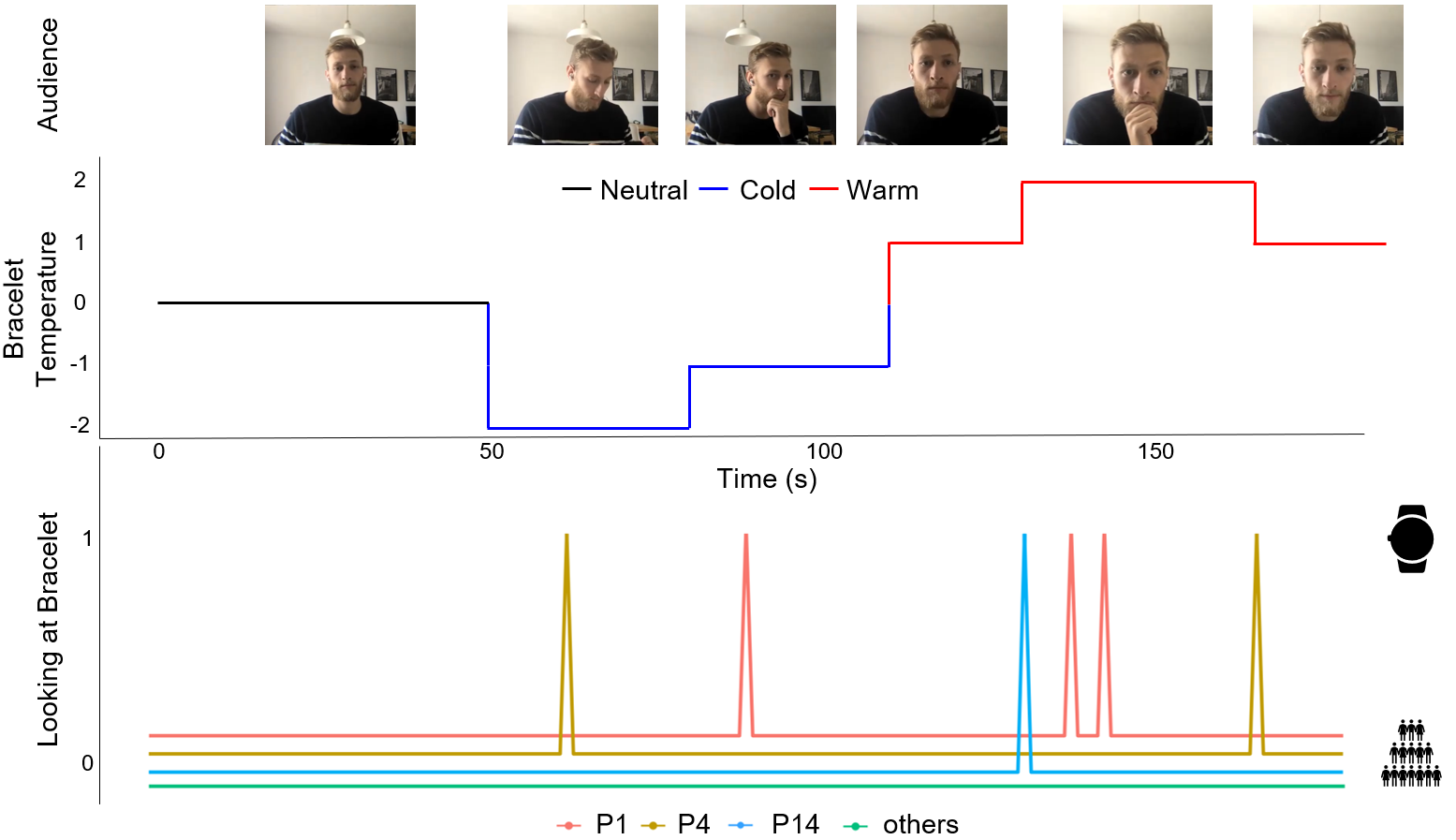}
    \caption{Overview of the experiment timeline. The plot shows example audience faces, the bracelet temperature level, and the times that participants looked at the thermal bracelet. The audience engagement follows a pattern of neutral, very disengaged, disengaged, engaged, very engaged, then engaged, with corresponding temperature changes}. All photos are used with the express written consent of the subjects.
     \Description{The figure shows an overview of the experiment timeline. The plot shows example audience engagement expressions, the normalized temperature level in the thermal feedback condition, and the times that participants looked at the thermal bracelet.}
     \label{fig:timeline}
\end{figure*}

We present both the qualitative and quantitative results clustered by the themes identified in the interview responses, first discussing the validity of the experiment and then providing evidence in the themes of distraction, social connectedness, and usability. Figure \ref{fig:timeline} shows an overview of the experimental procedure including a sample audience member, the thermal signal, and eye tracker information. 

\subsection{Validation}
Participants were consistent with their interpretation of the thermal signals. 90\% of the participants interpreted the signals as expected and associated an increased temperature with an engaged audience and a decreased temperature with a disengaged audience. Among those who chose the opposite mapping (i.e. set cold as engaged and warm as disengaged), P10 said that they dislike the feeling of warmth and disengagement is negative, thus assigning it the negative feeling of a disengaged audience. P4 mapped warmth to arousal and cold to focus.

The temperature range from $\pm$8\degree C to $\pm$8\degree C was comfortable for 17 of 20 participants. The coldest signal was uncomfortable for two participants, so they used a range of -6\degree C to 8\degree C, while one participant found the warmest signal uncomfortable and chose a range of -8\degree C to 6\degree C. When we presented the participants with the thermal feedback and asked them to recall their associated mapping, all 20 participants did so without errors.

Although the thermal mapping worked for communicating engagement, some participants found it counterproductive in the context of presentations; giving a presentation can be stressful~\cite{merz_oral_2019}, and stress is associated with heat~\cite{wilson_hot_2016}. For one participant, the thermal mapping to stress was more dominant than the mapping to engagement confirming the importance of personalization:
\begin{quote}
    ``\textit{If I get nervous or flustered during a talk [...] when it went hot [...] it was simulating that and it would make me maybe a bit more nervous [...]. The cold kind of did the opposite, it calmed me down a little bit.}'' (P17)
\end{quote}

The perceived stress was not normally distributed ($W=0.926$, $p=0.0121$), and Wilcoxon signed-rank tests with Bonferroni corrections indicate that there was no significant difference in stress between presenting with or without thermal feedback ($W=25$, $Z=-0.504$, $p=0.658$, $r=0.080$). Further, the workload according to the NASA-TLX was normally distributed ($W=0.983$, $p=0.788$) and showed no significant difference with or without thermal feedback according to a t-test ($t=-1.42$, $df=19$, $p=0.173$).

Participants' expected performance was not normally distributed ($W=0.925$, $p=0.0110$), so we performed Wilcoxon signed-rank tests with Bonferroni corrections. The participants did not expect to perform significantly better or worse during their talk with thermal feedback ($W=24$, $Z=-1.21$, $p=0.290$, $r=0.191$), which affirms that we did not artificially prime them with false confidence. There was also no significant difference in expected performance between topic conditions ($W=54$, $Z=-1.14$, $p=0.288$, $r=0.180$). Additionally, there were no significant correlations between the participants' average number of presentations per week and any of expected performance ($R=-0.072$, $p=0.657$), perceived connectedness ($R=-0.22$, $p=0.173$), or perceived quality ($R=-0.085$, $p=0.603$), so the results do not appear to be biased by prior presentation experience.

Regarding the feedback setup, we asked participants in the interview whether they would like the option to have additional channels representing individual members or segments of the audience. All of the participants indicated that the current setup, with a single aggregated channel representing the mean audience affect, was preferred as additional information ``\textit{would be more distracting}'' (P17).

\subsection{Distraction and Unobtrusiveness}\label{results:distraction}
All participants said they perceived the device and feedback to be unobtrusive and would not expect others to notice it. We asked each participant whether the audience would be able to notice the feedback: 85\%  noted that the signal itself is not detectable by others and 25\% mentioned that the device resembles a smartwatch so they would be comfortable wearing it during a real presentation. 

Although there was no significant increase in workload, seven participants voiced in the interview that they found the thermal feedback a bit distracting, and three quite distracting. Negative thermal feedback was mentioned as being particularly distracting because it caused participants to think about changing their talk.

Based on the eye tracking data, visualized in \autoref{fig:timeline}, only three participants looked at the thermal feedback device, one of them three times. They glanced at the device briefly but did not fixate for more than 400ms. Thus, the signal was not distracting to the point where participants were visually disrupted. The participants looked at the device mostly during very warm and very cold thermal feedback. The remaining participants did not find the thermal feedback distracting. Participants mentioned that they were able to detect the thermal feedback and understand its meaning without it interrupting them:

\begin{quote}
    ``\textit{I noticed it and I tried to adapt but it didn't bother me.}'' (P7)
\end{quote}




\begin{figure*}[ht]
 \centering
  \begin{subfigure}[b]{0.48\linewidth}
 \includegraphics[width=\linewidth, height=6cm]{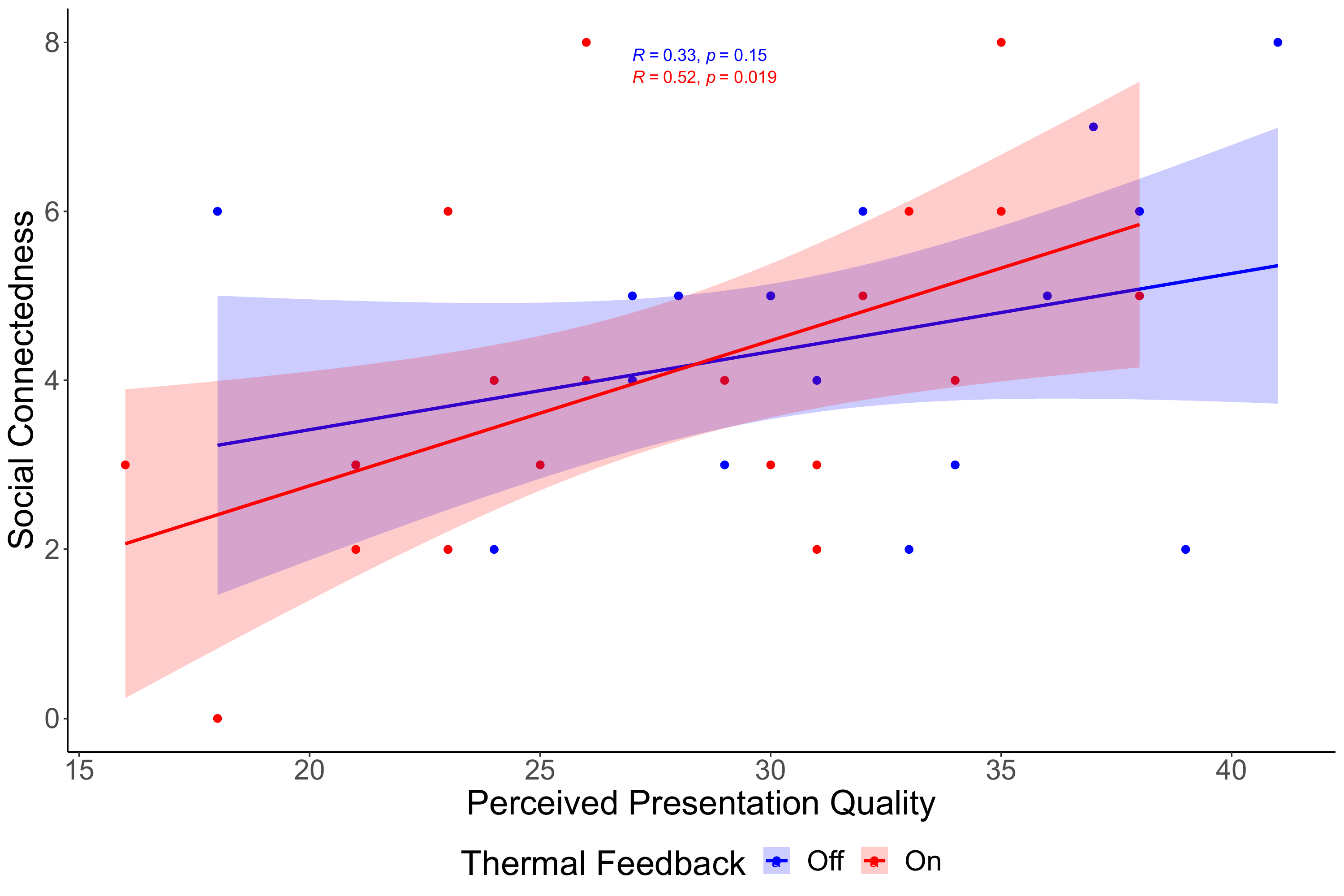}
 \subcaption{Social Connectedness and Perceived Presentation Quality. There is a significant positive Pearson correlation ($R=0.52$, $p=0.019$) only when using thermal feedback.}
 \label{fig:social_connection}
 \end{subfigure}
 \hspace{0.2em}
 \begin{subfigure}[b]{0.48\linewidth}
 \includegraphics[width=\linewidth, height=6cm]{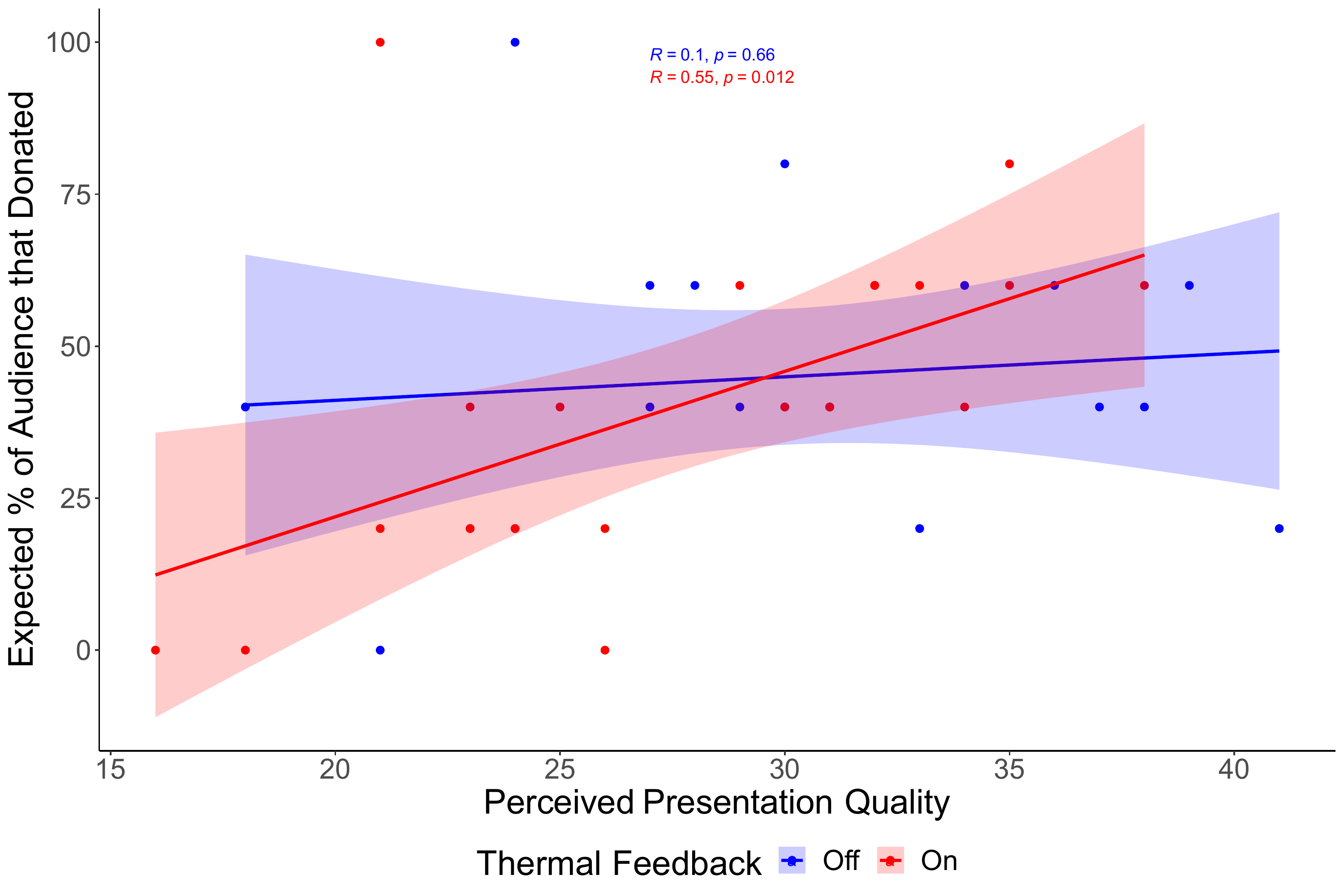}
 \subcaption{Donations and Perceived Presentation Quality. There is a significant positive Pearson correlation ($R=0.55$, $p=0.012$) only when using thermal feedback.}
 \label{fig:donations}
 \end{subfigure}
 \caption{Thermal feedback creates a significant correlation between Social Connectedness and Perceived Presentation Quality as well as Donations and Perceived Presentation Quality.}
 \Description{Two scatter plots showing the correlation between Social Connectedness and Perceived Presentation Quality as well as Donations and Perceived Presentation Quality. In both cases there is a significant correlation when using thermal feedback, and uncorrelated without.}
\end{figure*}

As a novel device and experience, participants said that the thermal feedback drew their attention. However, participants also expected that the thermal feedback would fade into the background with practice:

\begin{quote}
    ``\textit{At some point, you're going to still feel, but you're so used to it that you're not going to think about it anymore. It's like an intuition.}'' (P9)
\end{quote}

Participants found thermal feedback to be the most adequate modality they could imagine for communicating audience engagement, since other modalities would be more distracting: 
\begin{quote}
    ``\textit{I think nothing that is sound or too visual because that would put me off track for the presentation. What I liked about the temperature is that it's subtle.}'' (P18)
\end{quote}

\subsection{Social Connectedness}\label{section:socialconnection}
Questionnaire responses related to social connectedness varied depending on whether the participant was receiving positive or negative feedback. According to our results, participants felt reassured when they received positive thermal feedback and felt anxious when they received negative feedback. Participants also reported that, in general, the thermal feedback helped them feel more aware of and connected to the audience:

\begin{quote}
    ``\textit{Thermal feedback really did a great job in raising my awareness of the audience and also increases connectedness to my audience.}'' (P3)
\end{quote} 

Participants found the positive feedback reassuring and confidence inspiring:
\begin{quote}
    ``\textit{It was a bit calming because when it was warm, I knew that I was doing all right.}'' (P19)
\end{quote}


As mentioned in \autoref{results:distraction}, participants found negative feedback more distracting than positive feedback. One participant mentioned that they ``\textit{cannot really change anything at the moment}'' (P5), while others noted that the negative feedback was demotivating:

\begin{quote}
    ``\textit{If I'm already doing bad, I don't need to have someone tell me I'm doing bad. It's just demotivating.}'' (P19)
\end{quote}



Based on these responses, we hypothesized that participants would feel more positively about a good presentation with the thermal feedback than without, and more negatively about a bad presentation with thermal feedback than without. To test this hypothesis, we calculated Pearson's product-moment correlations for Social Connectedness and Perceived Presentation Quality. Social Connectedness, in this case, was measured by questions about how close or distant the participant felt from their audience. Further, we asked participants how many audience members they thought would donate to their simulated fundraiser and calculated Pearson's product-moment correlations for Donations and Perceived Presentation Quality.

As shown in \autoref{fig:social_connection}, Social Connectedness is significantly positively correlated ($R=0.52$, $p=0.019$) with Perceived Presentation Quality when the participants had thermal feedback. Without thermal feedback, there was no significant correlation ($R=0.33$, $p=0.15$).

Similarly, \autoref{fig:donations} shows that the expected number of Donations is significantly positively correlated ($R=0.55$, $p=0.012$) with Perceived Presentation Quality only when participants received thermal feedback ($R=0.1$, $p=0.66$ without thermal feedback).

\subsection{Usability and Contexts for Use}
\begin{figure*}[ht]
    \centering
    \includegraphics[width=\linewidth]{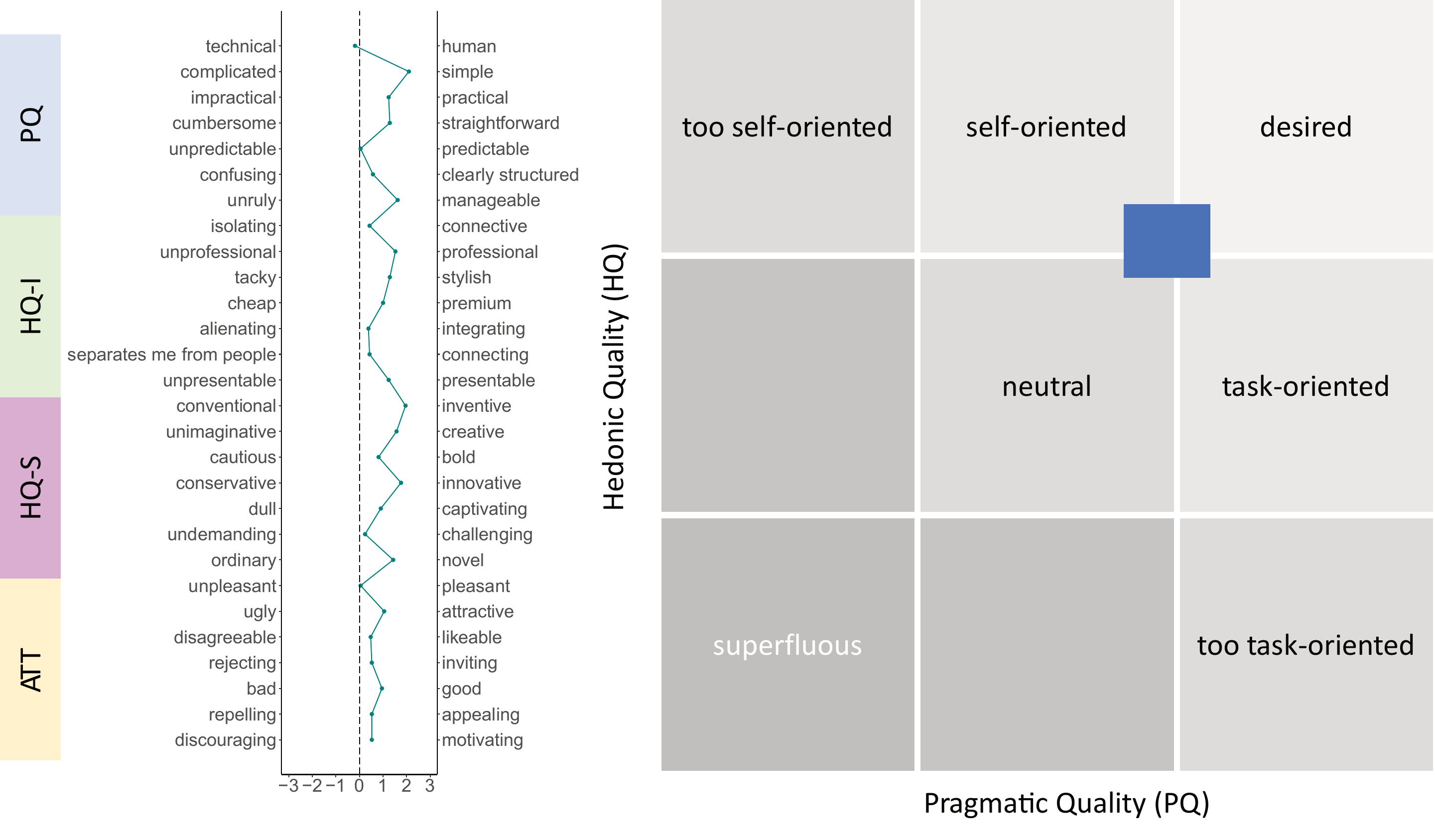}
    \caption{The mean results of the AttrakDiff questionnaire presented according to \cite{strockl_hedonic_2020}. The response categories are labeled Pragmatic Quality (PQ), Hedonic Quality Identity (HQ-I), Hedonic Quality Simulation (HQ-S) and Attractiveness (ATT). Hedonic Quality (HQ) = HQ-I + HQ-S.}
    \Description{The results of the AttrakDiff questionnaire. The response categories are labeled Pragmatic Quality (PQ), Hedonic Quality Identity (HQ-I), Hedonic Quality Simulation (HQ-S) and Attractiveness (ATT). Hedonic Quality (HQ) = HQ-I + HQ-S.}
     \label{fig:attrak_plot}
\end{figure*}

The results of the AttrakDiff questionnaire, shown in \autoref{fig:attrak_plot}, indicate that participants found the device to be usable. Aggregated participant responses indicate that overall thermal feedback was rated positively in terms of Pragmatic Quality (i.e. usefulness and efficiency), Hedonic Quality (i.e. joy of use), and Attractiveness. The confidence rectangle does not fall completely within any block, although it trends towards self-oriented, desired, neutral, and task oriented.

We asked all participants about the scenarios in which thermal feedback would be useful. Three participants mentioned that thermal feedback would be particularly useful for in-person scenarios because visual feedback would make them ``\textit{lose your eye contact}'' (P12), while in online presentations they are ``\textit{already looking at the screen}'' (P20). However, 40\% of participants felt that thermal feedback was best suited to online presentations and 35\% stated that it would be useful for both:
\begin{quote}
    ``\textit{I think it's probably more useful for online [presentations] because you don't get really much feedback at all. You can't hear people rustling in their chairs or anything like that.}'' (P17)
\end{quote}


Participants mentioned that for small in-person presentations or meetings the device would not be useful, since they would be able to understand audience engagement implicitly. However, participants mentioned that thermal feedback would be useful for in-person presentations in front of a large audience.

Finally, participants highlighted practice talks as a useful context. They stated that they would a similar system when giving a talk more than once so they could identify good and bad segments and iterate over their presentation:
\begin{quote}
    ``\textit{I think it would be really helpful if I'm practicing for a bigger presentations [...] I would practice and then the group of audience would give me feedback through this.}'' (P14)
\end{quote}

In this context, participants felt that the device could help a presenter ``\textit{to get better, and learn from your faults}'' (P15).

\section{Discussion}
\label{sec:discussion}
In this study, we sought to answer the research question: \textit{To what extent can thermal signals be used to communicate group affect and increase feelings of social connectedness over a distance?} We will therefore discuss how our findings relate to designing thermal systems for group settings, how thermal signals impact social connectedness, and implementing real-world audience feedback systems.

\subsection{Implications for Designing Thermal Feedback for Group Affect}
\subsubsection{Thermal Feedback Should Represent Dual Meanings at Most} Our study explored thermal feedback of group affect in the context of audience engagement. Although audience engagement can be complicated to measure~\cite{stevens_methods_2007}, the information that is being communicated to the presenter is simple. The thermal device communicates a two-sided signal, with one end being ``engaged'', the other being ``disengaged'', and no signal representing ``neutral''. This is in line with recommendations from both \citet{kappers_thermal_2019} and \citet{song_hot_2015} recommendation that temperature is not suitable to communicate complex messages. Based on our study as well as previous work, we recommend that future designers do not use a thermal signal to represent information that is more complex than a two-sided signal.

\subsubsection{Thermal Mapping Should be Personalized for Easy Understanding} It is crucial that the user understands and embodies the thermal mapping that is being communicated by the temperature signal in order to effectively use the system. Our results indicate that participants had no trouble internalizing the mapping that they chose for themselves, and although most agreed on the same mapping, there were two participants who chose the opposite. Our results are, therefore, in line with the results of Umair~\cite{umair_exploring_2021} and can underline that personalization is important for thermal displays for one-to-one and n-to-one communication in HCI.

\subsubsection{Negative Thermal Feedback Distracts While Positive Feedback Encourages}
When receiving negative feedback, participants tried to think of ways to change their presentation on the fly. In our study scenario, the participants had limited time to prepare their talk and had limited options for adapting their task based on audience feedback while talking. We expect that negative feedback could be effectively used if a presenter designed their presentation in advance with the intention of using an engagement-feedback system. However, based on the results of our study, we would recommend to design a feedback system that focuses on positive reinforcement if the goal is to improve social connectedness between the presenter and audience without distracting the presenter. Negative feedback should only be included in the system if the participants have the capacity to adapt to a negative signal, i.e., for proficient speakers, or to practice spontaneous adaptation. 

\subsection{Social Connectedness}
\subsubsection{Thermal Feedback Connects the Presenter to the Audience.}
When participants received thermal feedback, Social Connectedness and Perceived Presentation Quality as well as Donations and Perceived Presentation Quality covaried, while the same measures do not covary without thermal feedback. Although this does not necessarily indicate causation, the results show that the higher participants estimated their performance in their presentation, the more socially connected they felt while wearing the device. They also felt more distant from the audience when they did poorly only while wearing the device. The same holds true for Donations, which we used as a proxy for how well the presenter felt the audience would rate their presentation. This indicates that the connective function of the thermal feedback is not only present in participants qualitative statements but also increases awareness of group-communicative processes. Thus, enhancing the perception of group-affective states. It also highlights the suitability of the thermal signals to represent socio-emotional experiences, which is consistent with prior work showing that the temperature-emotion experience is contagious~\cite{cooper_you_2014} and that the socio-thermal mapping is tightly coupled~\cite{ijzerman_thermometer_2009}.

\subsubsection{Feedback is Useful for Disconnected Groups}
This paper targeted thermal feedback for communicating group affect, which differentiates from prior work that commonly explores one-to-one communication. Our participants highlighted that thermal feedback would be particularly useful for large audiences instead of more direct communication, since common non-verbal signals are easily accessible when presenting to a single person or a sufficiently small group. As such, the feedback from our participants motivates further research into the group context. The participants also mentioned that the feedback is well suited to communicating affect from online group members. It is more difficult to read non-verbal signals from online members, and at the extreme they may have their camera off and therefore give no opportunity for feedback. Past work has already investigated using thermal feedback to communicate between users at a distance through smartphones~\cite{wilson_thermal_2012, wilson_like_2011}, so these findings could be extended to synchronous virtual communication. The participants indicated that they were satisfied with a single aggregated channel that communicates the mean affect of the audience, but further research in other application scenarios may reveal that different communication strategies may be appropriate for certain tasks. For example, a teacher may be more concerned about outliers in their class (e.g., is there a student who is losing interest), rather than the average level of engagement, while an entertainer might be more concerned with maintaining a high average engagement.

\subsection{Towards Implementing a Real-World System}
Prior to this paper there was a multitude of research on measuring audience feedback (e.g.~\cite{rivera-pelayo_live_2013, murali_affectivespotlight_2021, hassib_engagemeter:_2017, martella_how_2015}, but very little on unobtrusively communicating this feedback to the presenter. Related prior work~\cite{costa_emotioncheck_2016} has also shown that wearable affective feedback devices can be effectively employed to impact affect with low attention requirements. By combining prior work on measuring audience engagement with our findings on providing feedback to presenters, all of the components required to implement a system in the real-world are now understood. Our system can scale to large audiences depending only on the limits of the chosen audience measuring system, so we could effectively transform what has been a one-to-one communication channel to be n-to-one.

Previous research in measuring and communicating audience engagement, such as ~\citet{murali_affectivespotlight_2021} and ~\citet{hassib_engagemeter:_2017}, commonly used visual methods, such as charts and bars, to represent the affective information from the audience. Future work should investigate how thermal and visual representations could be combined. For example, the discrete nature of thermal signals could be used during a presentation, while a detailed visual representation could be provided afterwards to encourage reflection and improvement over time.

\subsubsection{Users Expect Our Thermal Feedback System to Unobtrusively Support Their Main Task in the Real-World}
Our study used a Wizard of Oz approach to generate a thermal signal rather than measuring actual audience engagement. To employ a socio-thermal feedback system in the wild, our feedback setup should be combined with prior work in measuring audience engagement, such as the facial expression recognition system from~\citet{murali_affectivespotlight_2021} for a virtual audience or the EngageMeter from~\citet{hassib_engagemeter:_2017} for an in-person audience. Our participants reported that the device could be worn in a real group scenario without being detected, and they expected the signal to fade into the background over time. Moving technology into the periphery of attention is the primary focus of unobtrusive design~\cite{bakker_design_2013, weiser_coming_1997}, so we expect that our system can be implemented in the real world unobtrusively.

\subsubsection{Combining Thermal Feedback with Implicit Audience Measurement May Facilitate Non-verbal Remote Communication}
According to the design space proposed by Hassib et al.~\cite{hassib_design_2018}, an implicit and synchronous feedback system would be most appropriate to combine with thermal feedback since it supplies an embodied continuous signal. As mentioned in the previous section, our system could be combined with implicit audience measurement systems such as ~\citet{murali_affectivespotlight_2021} for virtual audiences or~\citet{hassib_engagemeter:_2017} for in-person scenarios. The combination of thermal feedback with implicit audience measurement has the potential to overcome the lack of non-verbal cues associated with online and hybrid presentation settings~\cite{saatci_hybrid_2019} by providing a feasible, unobtrusive system that enhances social connectedness between presenters and their audience without increasing workload.

\subsection{Limitations \& Future Work}
One potential limitation of our study is the influence of the novelty effect~\cite{kormi-nouri_novelty_2005}. The participants were not accustomed to interacting with thermal signals, so the device was a novel experience. If the participants were to use the device over time, the signal would likely become more implicit and less distracting.

Another limitation of our study is that we used an eye tracker to identify moments where participants were distracted to the point that they looked at the bracelet. These disruption events likely do not fully describe the level of distraction that the participants experienced. However, we captured more detailed subjective responses regarding distraction in the interviews, where participants reported that negative feedback was distracting. Future work could characterize the level of distraction caused by the thermal device in more detail using, for example, EEG~\cite{thiruchselvam_temporal_2011}.

Additionally, the use of a recorded audience can be considered a limitation of our study. Since the audience was pre-recorded, their reactions were not necessarily aligned with the quality of each presentation. However, the recorded audience was a more controlled condition than a live audience. Since our study is the first to explore the communication of social signals with temperature in a group communication setting, we decided that the controlled condition was more appropriate. Also, the focus of our study was to explore the display of group-affective states, rather than the measurement of those states. Still, there is an opportunity for future research to implement our system in the wild to gain further understanding of how thermal feedback could be employed in real group communication. In such an in-the-wild experiment, it would also be interesting to investigate different methods of audience feedback, such as implicit sensing or periodic explicit responses, to see which paradigm best integrates with the thermal feedback.

Finally, the inability of participants to adapt their talks can be considered a limitation. Providing all participants with the same two topics established controlled experimental conditions, but future work may benefit from letting participants present topics they are familiar with, allowing them to adapt their talks. Exploring longer-format presentations, such as teaching, could also give participants more opportunity to adapt to the thermal feedback they receive. 

\section{Conclusion}
Our research and the presented study provide the first experimental exploration of thermal feedback to communicate group affect in the context of hybrid presentations. We contribute empirically validated knowledge that informs designers of social-emotional feedback systems that use temperature as a modality. Our findings indicate that temperature can be an effective and usable way to communicate audience engagement to a presenter and increase their feeling of connection to the audience without increasing workload. Our research shows that personalization of the thermal feedback is crucial and that users can quickly learn an established mapping and to effectively use the system. In general, participants found positive (warm) feedback to be encouraging and useful while (cold) negative feedback was seen as distracting. Our findings lay the foundation for engineering thermal feedback systems and show that such systems have the potential to enhance socio-emotional connectedness in virtual and hybrid meetings, where users currently feel disconnected. Furthermore, our work is valuable for practitioners and researchers who aim to implicitly increase social connectedness through technology.

\begin{acks}
This work was supported by the Bavarian Research Alliance association ForDigitHealth, by the European Union’s Horizon 2020 Programme under ERCEA grant no. 683008 AMPLIFY, and by the Munich Center for Machine Learning (MCML). We also want to thank Matt Smith from Embr Labs for showing an interest in our research and providing the Embr Wave bracelets for our study.
\end{acks}

\bibliographystyle{ACM-Reference-Format}
\bibliography{references}

\end{document}